\documentclass[pra,twocolumn,a4paper,twoside,showkeys,showpacs]{revtex4}
\usepackage{times,amssymb,graphicx,bbm,amsmath}

\newcommand{\be}{\begin{equation}}
\newcommand{\ee}{\end{equation}}
\newcommand{\bea}{\vspace{0.25cm}\begin{eqnarray}}
\newcommand{\eea}{\end{eqnarray}}
\begin{document}
\title{Measuring the photon distribution by ON/OFF photodectors}
\author{Marco Genovese, Marco Gramegna, Giorgio Brida}
\affiliation{Istituto Elettrotecnico Nazionale, IEN, Galileo Ferraris, Torino, Italia}
\author{Maria Bondani, Guido Zambra, Alessandra Andreoni}
\affiliation{INFM-CNR and Dipartimento di Fisica e Matematica,
Universit\`a degli Studi dell'Insubria, Como, Italia}
\author{Andrea R. Rossi, Matteo G. A. Paris}
\affiliation{Dipartimento di Fisica dell' Universit\`a di Milano, Italia.}
\begin{abstract}
Reconstruction of photon statistics of optical states
provide fundamental information on the nature of any optical field
and find various relevant applications. Nevertheless, no detector
that can reliably discriminate the number of incident photons is
available. On the other hand the alternative of reconstructing
density matrix by quantum tomography leads to various technical
difficulties that are particular severe in the pulsed regime
(where mode matching between signal an local oscillator is very
challenging). Even if on/off detectors, as usual avalanche
PhotoDiodes operating in Geiger mode, seem useless as photo
counters, recently it was shown how reconstruction of photon
statistics is possible by considering a variable quantum
efficiency. Here we present experimental reconstructions of photon
number distributions of both continuous-wave and pulsed light
beams in a scheme based on on/off avalanche photodetection
assisted by maximum-likelihood estimation. Reconstructions of the
distribution for both semiclassical and quantum states of light
(as single photon, coherent, pseudothermal and multithermal
states) are reported for single-mode as well as for multimode
beams. The stability and good accuracy obtained in the
reconstruction of these states clearly demonstrate the interesting
potentialities of this simple technique.
\end{abstract}
\date{\today}
\maketitle
\section{Introduction}
The evaluation of diagonal elements of the density matrix for
quantum optical states, i.e. of the statistical distribution of
the number of photons, provides fundamental information on the
nature of any optical field and finds various relevant
applications, ranging from studies on foundations of quantum
mechanics \cite{PR} to quantum information \cite{NC}, and quantum
metrology. Despite the importance of photon distribution, photon
detectors allowing an effective discrimination among different
number of incident photons are not yet available. Among the
possible candidates to number resolving photo-detectors,
PhotoMultiplier Tubes (PMT's) \cite{burle} and hybrid
photodetectors \cite{NIST} have the drawback of a low quantum
efficiency, since the detection starts with the emission of an
electron from the photocathode. On the other hand, solid state
detectors with internal gain, in which the nature of the primary
detection process ensures higher efficiency, are still under
development. Highly efficient thermal photon counters have also
been used. However, since they operate in cryogenic conditions
they are far to allow common use \cite{xxx,serg}. Furthermore,
their efficiency is limited by the optical window for entering the
cryostat.
\par
Quantum tomography provides an alternative method to measure
photon number distributions \cite{mun}. However, the tomography of
a state, which has been applied to several quantum states
\cite{raymerLNP}, needs the implementation of homodyne detection,
which in turn requires the appropriate mode matching of the signal
with a suitable local oscillator at a beam splitter. This
technique is therefore, in general, not of simple implementation
and, in particular, such mode matching is a very challenging task
in the case of pulsed optical fields.
\par
On the other hand, the photodetectors usually employed in quantum
optics, such as Avalanche PhotoDiodes (APD's) operating in the
Geiger mode \cite{rev,serg} (that have relatively large quantum
efficiencies), are not suited for distinguishing different number
of incident photons, since they have the obvious drawback that the
breakdown current is independent of the number of detected
photons. The outcome of these APD's is either "off" (no photons
detected) or "on" {\em i.e.} a "click", indicating the detection
of one or more photons. Actually, such an outcome can be provided
by any photodetector (PMT, hybrid photodetector, cryogenic thermal
detector) for which the charge contained in dark pulses is
definitely below that of the output current pulses corresponding
to the detection of at least one photon. Notice that for most
high-gain PMT's the anodic pulses corresponding to the
"no-photons" ("no-click") event can be easily discriminated by a
threshold from those corresponding to the detection of one or more
photons. On the other hand, , as we will describe in the next
paragraph, these detectors can be used for reconstructing photon
statistic when measurements at different quantum efficiencies are
performed.
\par
In this paper we present in some details (see \cite{nos} for a
shorter summary of these results) experimental reconstructions of
photon number distributions of both continuous-wave and pulsed
light beams in a scheme based on on/off avalanche photodetection
assisted by maximum-likelihood estimation. Reconstructions of the
distribution for both semiclassical and quantum states of light
(as single photon, coherent, pseudothermal and multithermal
states) are reported for single-mode as well as for multimode
beams.
\section{Theoretical method}
The statistics of the "no-click" and "click" events from an on/off
detector, assuming no dark counts, is given by
\begin{eqnarray}
p_0(\eta) &=& \sum_n (1-\eta)^n \varrho_n\: \label{p0}\:,
\end{eqnarray}
and $p_{>0} (\eta) =1- p_0(\eta)$, where $\varrho_n =\langle n |
\varrho | n \rangle$ is the photon distribution of the quantum
state $\varrho$ and $\eta$ is the quantum efficiency of the
detector, {\em i.e.} the probability of a single photon to be
revealed.  At first sight the statistics of an on/off detector
appears to provide quite a scarce piece of information about the
state under investigation. However, if the statistics about
$p_0(\eta)$ is collected for a suitably large set of efficiency
values then the information is enough to reconstruct the whole
photon distribution $\varrho_n$ of the signal, upon a suitable
truncation at $\bar{n}$ of the Hilbert space.
\par
The reconstruction of photon distribution through on/off detection
at different efficiencies has been analyzed \cite{mogy} and its
statistical reliability investigated in some details
\cite{pcount}. In addition, the case of few and small values of
$\eta$ \cite{ar} has been addressed. However, whilst these
theoretical studies found an application to realize a multichannel
fiber loop detector \cite{olom,kb}, an experimental implementation
of this technique for reconstructing photon distribution of a
free-propagating field is still missing. In view of the relevance
of photon distribution for applications in quantum information and
foundations of quantum mechanics, our purpose is to show that a
reconstruction of the photon distribution by using this technique
can be effectively realized gathering results obtained from
measurement at different quantum efficiencies. As we will see this
method leads to excellent results both for free-propagating
continuous-wave (cw) and pulsed light beams, for both single-mode
semiclassical and quantum states, as well as for multimode states.
\par
The procedure consists in measuring a given signal by on/off
detection using different values $\eta_\nu$ ($\nu=1,...,K$) of the
quantum efficiency.  The information provided by experimental data
is contained in the collection of frequencies $f_{\nu} =
f_0(\eta_\nu) = n_{0\nu}/n_\nu$ where $n_{0\nu}$ is the number of
"no click" events and $n_\nu$ the total number of runs with
quantum efficiency $\eta_\nu$.  Then we consider expression
(\ref{p0}) as a statistical model for the parameters ${\varrho_n}$
to be solved by maximum-likelihood (ML) estimation. Upon defining
$p_\nu\equiv p_0(\eta_\nu)$ and $A_{\nu n} = (1-\eta_{\nu})^n$ we
rewrite expression (\ref{p0}) as $p_{\nu} = \sum_{n} A_{\nu n}
\varrho_n$. Since the model is linear and the parameters to be
estimates are positive (LINPOS problem), then the solution can be
well approximated by using the Expectation-Maximization algorithm (EM)
\cite{EMalg}. By imposing the restriction $\sum_n \varrho_n = 1$,
we obtain the iterative solution
\begin{equation}
\varrho_n^{(i+1)} = \varrho_n^{(i)}\sum_{\nu=1}^K \frac{A_{\nu
n}}{\sum_\lambda A_{\lambda n}}
\frac{f_{\nu}}{p_{\nu}[\{\varrho_n^{(i)}\}]}\: \label{ems}\:
\end{equation}
where $p_{\nu}[\{\varrho_n^{(i)}\}]$ are the probabilities
$p_{\nu}$, as calculated by using the reconstructed distribution
$\{\varrho_n^{(i)}\}$ at the $i$-th iteration. \\
\begin{figure}[h]
\includegraphics[width=0.3\textwidth]{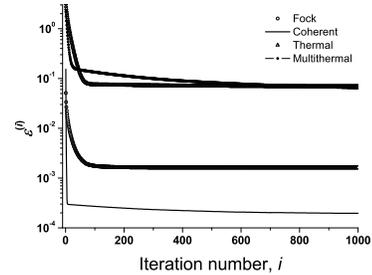}
\caption{\label{f:convergence} Plot of the total error
$\varepsilon^{(i)}$ as a function of the iteration number for the
various state of the field (corresponding to the states
experimentally investigated, see next Sections). For the thermal
distribution the value of the total error is plotted at each
iteration step (value on the $x$ axis), while for the other states
the value of $\varepsilon^{(i)}$ is the average over an suitable
set of iterations.}
\end{figure}
Before going to the experimental implementation we have performed
several numerical simulations in order to check the accuracy and
reliability of this method by varying the different parameters.
Since the solution of the ML estimation is obtained iteratively,
the most important aspect to keep under control is its
convergence. Of course, the degree of convergence at a given step
can be checked evaluating the log-likelihood function
$\mathcal{L}=\log{L}$, $$L = \prod_{\nu=1}^{K}
p_\nu^{n_{0\nu}}\left( 1-p_\nu\right)^{n_\nu-n_{0\nu}}\:.$$
However, a more suitable parameter is given by the total absolute
error at the $i$-th iteration {\em i.e.}
\begin{equation}
\varepsilon^{(i)} =\sum_{\nu=0}^K \left| f_\nu- p_\nu [\{
\varrho_n^{(i)}\}]\right|\:.
\end{equation}
Indeed,  the total error measures the distance of the
probabilities $p_\nu [\{ \varrho_n^{(i)}\}]$, as calculated at the
$i$-th iteration, from the actual experimental frequencies and
thus, besides convergence, it quantifies how the estimated
distribution reproduces the experimental data. The total distance
is a decreasing function of the number of iterations. Its
stationary value is proportional to the accuracy of the
experimental frequencies $\{f_\nu\}$.  For finite data sample this
value is of order $1/\sqrt{n_\nu}$ for each value of $\eta_\nu$,
giving us a rough estimate of $\bar{n}/\sqrt{n_\nu}$ for the total
error for the reconstructed probability $p_\nu [\{
\varrho_n^{(i)}\}]$. If the stationary value of
$\varepsilon^{(i)}$ is of this order we have double checked the
convergence of the whole method. Notice that these properties are
not shared by the log-likelihood: its stationarity certainly
reveals convergence, but the value depends on the statistics to be
retrieved and so can not even be estimated by \emph{a priori}
considerations.
\par
In order to check the convergence of the iterations in
(\ref{ems}), we run simulated experiments to reconstruct some of
the states subsequently investigated experimentally (see next
Sections and Figs. \ref{f:fock}-\ref{f:total}). The results are
shown in Fig. \ref{f:convergence}, where the total error (ratio
its stationary value) is reported as a function of the number of
iterations. As it is apparent from the plot the total error shows
a transient behaviour and then quickly converges to its stationary
value. The rate of convergence depends on the signal under
investigation.  Our results show that the iterative algorithm
always converges and the asymptotic value of $\varepsilon^{(i)}$
is of the expected order on magnitude.
\par
An estimate of the confidence interval on the determination of the
element $\varrho_n$ can be given in terms of the variance
\begin{eqnarray} \label{confidence}
\sigma_n = (\mathcal{N}\,F_{n})^{-1/2}\,,
\end{eqnarray}
$\mathcal{N}$ being the total number of measurements, and $F_{n}$
the Fisher's information \cite{cramer}
\begin{equation}\label{fisher}
F_n=\sum_{\nu} \frac{1}{q_{\nu}} \left( \frac{\partial
q_{\nu}}{\partial \varrho_n} \right)^2\:,
\end{equation}
where
\begin{equation}
q_\nu = \frac{p_\nu}{\sum_\nu p_\nu} = \frac{\sum_n A_{\nu
n}\varrho_n}{ \sum_{\nu n} A_{\nu n} \varrho_n}\:,
\end{equation} represents the renormalized probabilities of the
{\em no-click} event at quantum efficiency $\eta_\nu$.
\section{Experimental data in cw regime}
In the following we present various different applications of this
method both to cw and pulsed regime, with the purpose of
demonstrating the potentialities of this technique. For what
concerns cw regime we have studied reconstruction of diagonal
element of the density matrix for single photon Fock states and a
weak coherent one.
\par
\begin{figure}[h]
\includegraphics[width=0.3\textwidth]{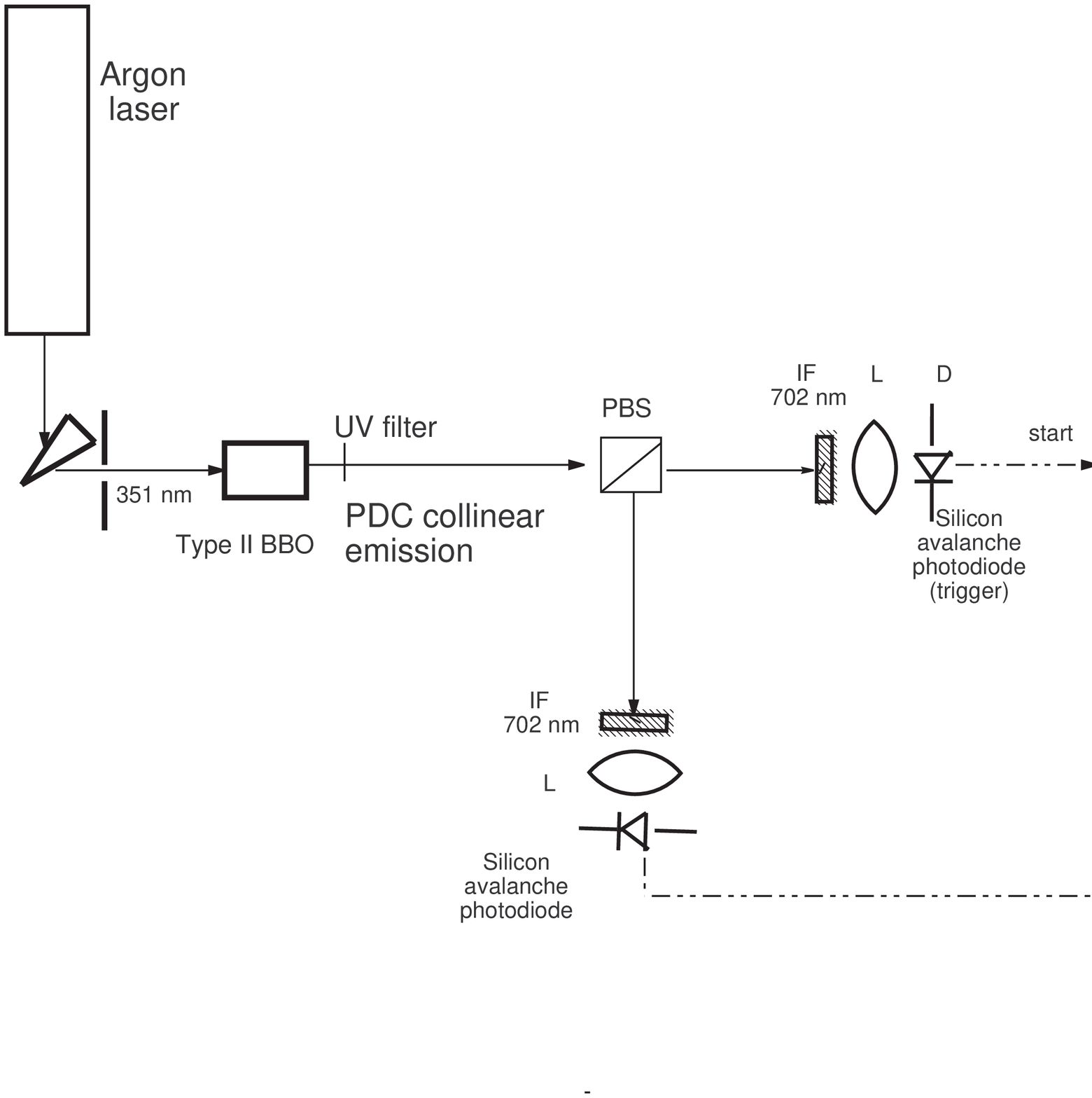}
\caption{\label{f:setup} Set up for heralded single PDC photon
reconstruction. An Ar$^{+}$ laser beam pumps a type II BBO crystal
generating collinear degenerate PDC. After an anti-UV filter,
pairs are split on a PBS. Observation of a photon on trigger
detector D (preceded by a lens, L, and an interference filter IF)
starts a TAC ramp which is eventually closed by a stop signal
deriving from the observation of a photon in the second detector.}
\end{figure}
The single photon states have been generated by producing
Parametric Down Conversion (PDC) heralded photons. In some more
detail (see Fig. \ref{f:setup}), pairs of correlated photons have
been generated by pumping in collinear degenerate geometry a 5 mm
long $\beta$-Barium-Borate (BBO) crystal, cut for type II phase
matching (i.e the two photons of the pair have orthogonal
polarizations), with a 0.3 W cw Argon ion laser beam with
wavelength of 351 nm. The heralded single photon scheme is based
on the specific properties of PDC emission. PDC  is a quantum
effect without classical counterparts and consists of a
spontaneous decay, inside a non-linear crystal, of one photon from
a pump beam (usually generated by a laser) into a couple of
photons conventionally called signal and idler. This decay process
obeys (phase matching laws) to energy conservation \be \omega_0 =
\omega_i + \omega _s \ee
 and momentum conservation \be \vec{k}_0 = \vec{k}_i + \vec{k} _s \ee where
$\omega_0,  \omega_i,  \omega _s$ are the frequencies and
$\vec{k}_0,  \vec{k}_i,  \vec{k} _s$ the wave vectors of pump,
idler and signal photon respectively.
 Furthermore, the two
photons are produced, within few tens femtoseconds, at the same
time. The probability of a spontaneous decay into a pair of
correlated photons is usually very low, of the order of $10^{-9}$
or lower; therefore with typical pump power of the order of some
milliwatts, the fluorescence emission lies at the levels of photon
counting regime. Since the photons are produced in pairs and
because of the energy and momentum conservation restrictions, the
detection of one photon in a certain direction and with a given
energy indicates the existence of the pair correlated one, with
definite energy in a well defined direction. This property allows
to "heralding" the second photon of the pair once the first one is
detected in a precise direction (and temporal and spectral
window). This "heralded photon" was the state to be measured. In
our set-up, after having eliminated the pump laser beam with a
filter, the two photons of the pair are separated by means of a
polarizing beam splitter. The detection of one photon in one of
the two conjugated directions  was then used to open a window of
$\Delta t =4.9 ns$ for detection in arm 2. This was realised by
addressing the first detection signal as a start to a Time to
Amplitude Converter, the signal from second detector (after a
delay line) was then addressed to the same TAC as a stop and
counted only if arriving in a window of $\Delta t$.
\par
\begin{figure}[h]
\includegraphics[width=0.25\textwidth]{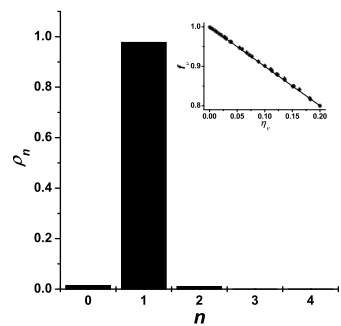}
\caption{\label{f:fock} Reconstruction of the  photon distribution
for the heralded single-photon state produced in spontaneous type
II PDC. Inset: Experimental frequencies $f_{\nu}$ of no-click
events as a function of the quantum efficiency $\eta_{\nu}$ for a
PDC heralded photon state compared with the theoretical curve
$p_\nu = 1-\eta_\nu$ .} \end{figure}
 The photodetection
apparatuses were constituted by  a silicon avalanche photodiode
detector preceded by an iris and an interference filter (IF) at
702 nm, 4 nm FWHM, inserted with the purpose of reducing the stray
light. Both the detectors were silicon avalanche photodiode ones
(SPCM-AQR-15, Perkin Elmer). The quantum efficiency of the
"heralded photon" detection apparatus (including IF and iris) was
measured to be $20 \%$ by using the standard calibration scheme
based on correlation properties of PDC emission (see
\cite{pdccs}). Lower quantum efficiencies, needed for the
reconstruction scheme, were simulated by inserting calibrated
neutral density filters on the optical path. A comparison of the
observed frequencies $f_{\nu}$ with the theoretical curve ($1-
\eta_{\nu}$) is presented in the inset of Fig. \ref{f:fock}.  The
photon distribution has been reconstructed using $K=34$ different
values of the quantum efficiency from $\eta_{\nu}  \simeq 0.01 \%$
to $\eta_\nu\simeq 20\%$ with $n_\nu=10^6$ runs for each
$\eta_\nu$. Results at iteration $i=10^6$ are shown in Fig.
\ref{f:fock}. As expected the PDC heralded  photon state largely
agrees with a single photon Fock state. However, also a small two
photons component and a vacuum one are observed.  The $\rho_2$
contribution is expected, by estimating the probability that a
second photon randomly enters the detection window, to be $1.85
\%$ of $\rho_1$, in agreement with what observed. A non zero
$\rho_0$ is also expected due to background.  This quantity can be
evaluated  by measuring the counts when the polarization of the
pump beam is rotated by a $\lambda / 2$ wave plate to avoid
generation of parametric fluorescence. In this case as well, our
estimate, $(2.7 \pm 0.2) \%$, is in good agreement with the
reconstructed $\rho_0$.
\par
\begin{figure}[h]
\includegraphics[width=0.25\textwidth]{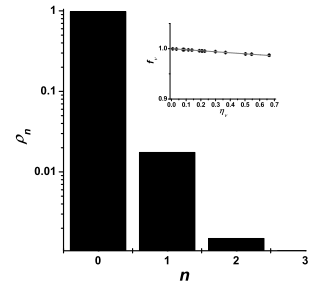}
\caption{\label{f:weakco} Reconstruction of the  photon
distribution for a weak coherent state. In the inset the
experimental frequencies $f_{\nu}$ of no-click events as a
function of the quantum efficiency $\eta_{\nu}$ for a PDC heralded
photon state are compared with a typical curve for a weak coherent
beam $p_\nu \simeq 1-\eta_\nu |\alpha|^2$.}
\end{figure}

As a second example we have reconstructed the statistic of a
strongly attenuated coherent state, which has been produced by a
He-Ne laser beam attenuated to photon-counting regime  by
insertion of neutral filters.  The same silicon avalanche
photodiode detector of the previous case was used here as well.
The counts were measured in about 400 ns window obtained by gating
the photo-detector with a periodic signal (10 kHz rate). It must
be noticed that in this case we do not have interference filters
or irises in front of the detector and all the other attenuations
can be included in the generation of the state (i.e. they
contribute to the absorbtion together with neutral filters): thus
the highest quantum efficiency is assumed to be $66 \%$ as
declared by the manufacturer data-sheet for the photodetector. The
reconstructed distribution, with $K=15$ different values of the
quantum efficiency from $\eta_{\nu}  \simeq 0.1 \%$  to
$\eta_\nu\simeq 66\%$ with $n_\nu=10^6$ runs for each $\eta_\nu$,
agrees well with what expected for a coherent state with average
number of photons $|\alpha|^2 \simeq 0.02$. In  Fig.
\ref{f:weakco} are shown both the frequencies $f_{\nu}$ as a
function of $\eta_{\nu}$  compared with the theoretical prediction
$p_\nu = \exp\{-\eta_\nu |\alpha|^2\} \simeq 1- \eta_\nu
|\alpha|^2$ and the reconstructed photon statistic. Finally, we
briefly acknowledge that a comparable result was also obtained
with a very strongly attenuated thermal state, i.e. light emitted
by a thermal source, a tungsten lamp attenuated by neutral density
filters.
\section{Experimental data in pulsed regime}
\begin{figure}[h]
\includegraphics[width=0.15\textwidth,angle=-90]{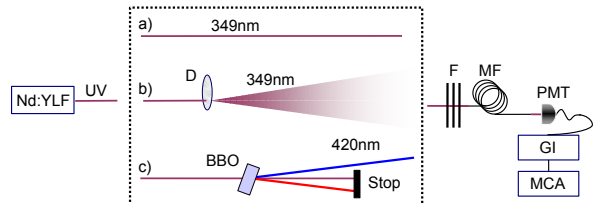}
\caption{\label{f:setupPULSED} Setup for the generation of pulsed
optical states. Starting with the third-harmonics of a Nd:YLF
laser, we obtain: a) gaussian state; b) thermal state; c)
multithermal state. D, rotating ground-glass diffuser; BBO, type I
nonlinear crystal; F, variable filter; MF, multimode fiber; PMT,
photomultiplier detector; GI, gated integrator; MCA, multichannel
analyzer.}
\end{figure}

\begin{figure}[h]
\includegraphics[width=0.3\textwidth]{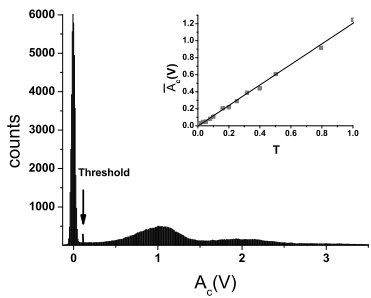}
\caption{\label{f:threshold} Histogram of the detected anodic
charge of the PMT (data corresponding to the measurement of a
gaussian field having $N=4.88$ with $\eta = 0.20$). The arrow
indicates the threshold value used to discriminate on/off events.
Inset: linearity test for the PMT detector: mean anodic current as
a function of the transmittance of calibrated filters.}
\end{figure}

In the pulsed domain, we have measured three different optical
states generated starting from the third harmonics (349 nm, 4.45
ps) of a cw mode-locked Nd:YLF laser regeneratively amplified at a
repetition rate of 500 Hz (High Q Laser). The general experimental
setup is sketched in Fig.~\ref{f:setupPULSED}. For all the
measurements, the light was delivered to a photo multiplier tube
(PMT, Burle 8850) through a multi mode fiber (100 $\mu$m core
diameter). Although the PMT has the capability of counting the
number of photoelectrons produced by one or more photons
\cite{burle}, for the present application we used it in a
Geiger-like configuration, by setting a threshold to discriminate
on/off events. Furthermore we take advantage of the linearity of
the mean anodic charge, $\overline{A^c}$, see Inset in
Fig.~\ref{f:threshold}, as a function of the mean energy of the
measured light \cite{PMTexamp} to obtain the values of $\eta_\nu$.
In fact, if we set $\overline{A^c}=0$ for $\eta=0$ and
$\overline{A^c}=(\overline{A^c})_{\mathrm{max}}$ for
$\eta=\eta_{P}$ (nominal quantum efficiency of the PMT), for all
the intermediate quantum efficiencies obtained by attenuating the
light with neutral filters, we have
$\eta=(\eta_{P}/(\overline{A^c})_{\mathrm{max}})\overline{A^c}$.
This procedure allow us to vary continuously the quantum
efficiency.
\par
\begin{figure}[h]
\includegraphics[width=0.25\textwidth]{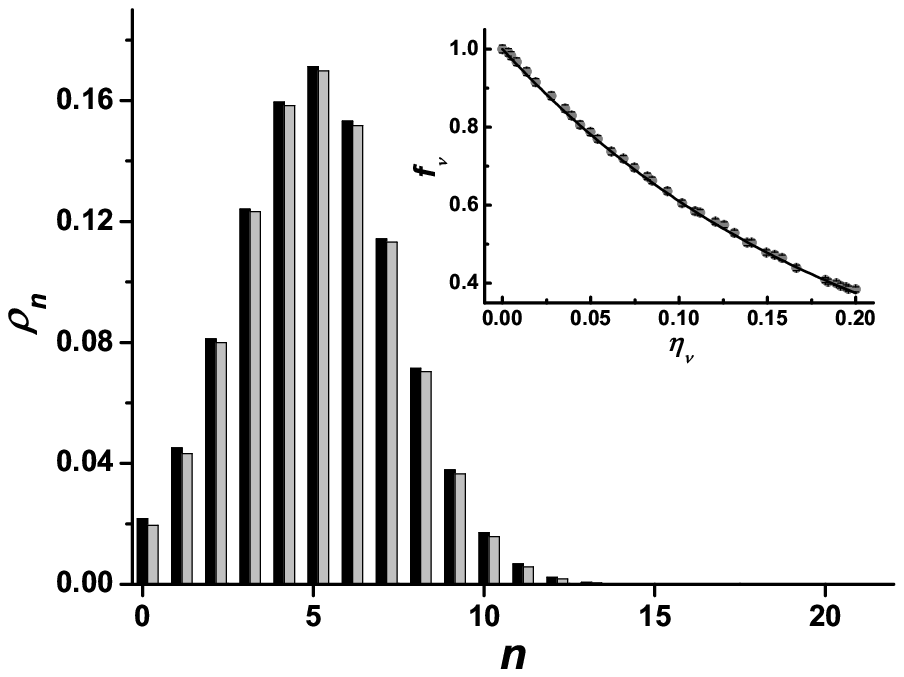}
\includegraphics[width=0.25\textwidth]{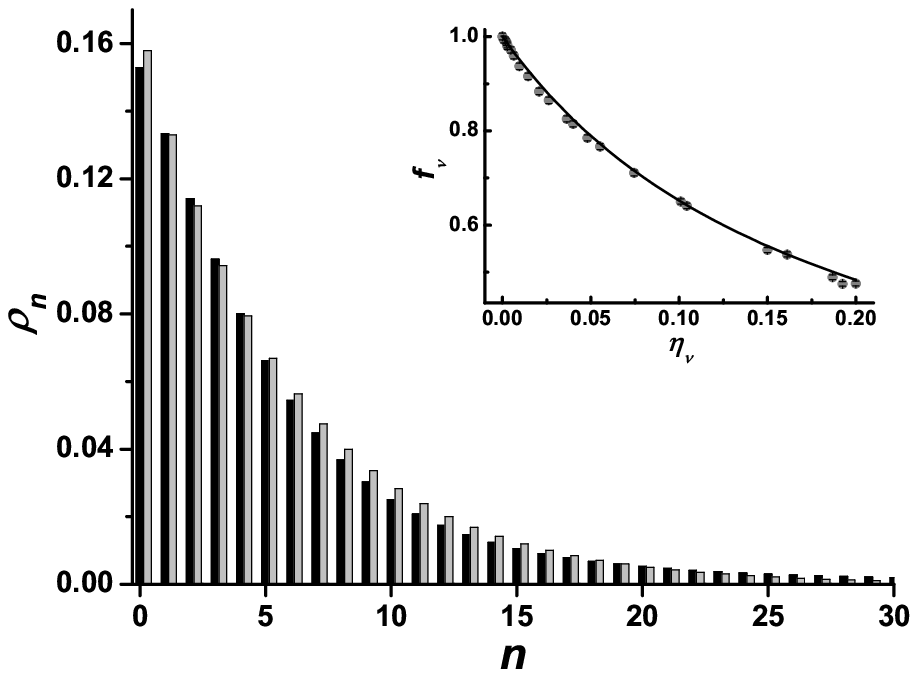}
\includegraphics[width=0.25\textwidth]{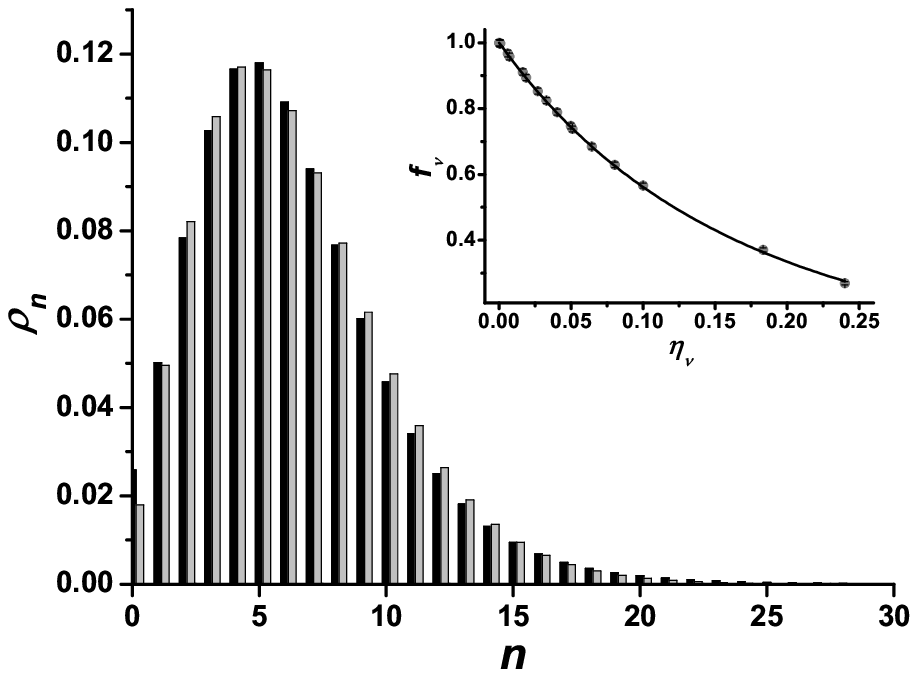}
\caption{Reconstructed photon distribution (black bars) and best
theoretical fit (grey bars) for three different states in the
pulsed regime: a) laser pulse, b) diffused laser pulse, c)
multimode state produced in type I PDC. Insets: Experimental
frequency $f_{\nu}$ data in function of the quantum efficiency
$\eta_{\nu}$ and theoretical model for each one of the
states.\label{f:total}}
\end{figure}

The first measurement was performed on the pulse emerging from the
laser source. Due to the pulsed nature of the source, we do not
expect to recover a true Poissonian statistics as in the cw
measurement described above. Rather, we expect a Gaussian
distribution of the form \cite{lasergaus}
\begin{equation}
\varrho_{n,G} = \frac{1}{\sqrt{2\pi (N+\sigma^2)}} \exp
\left[-\frac{(n-N)^2}{2(N+\sigma^2)}\right]\:, \label{gauss}
\end{equation}
which takes into account the presence of noise; $N$ is the photon
mean value and $\sigma^2/N$ can be taken as a measure of the
deviation from Poissonian statistics. In Fig. \ref{f:total} a) we
show the photon distribution $\varrho_n$, reconstructed at the
$i=50000$ iteration of the ML algorithm, along with the best fit
obtained with the model (\ref{gauss}) (fitting parameters $N=
4.88$ and $\sigma^2 = 0.63$). The inset of the figure compares the
experimental frequency $f_{\nu}$ data ($K=37$ values of $\eta$,
$n_\nu=10^4$ runs for each $\eta$) as a function of $\eta_\nu$
with the theoretical values calculated through (\ref{p0}) and the
parameters given by the fit of the photon distribution. Both the
reconstructed distribution and the experimental frequencies agrees
very well with the above Gaussian model. The fidelity of the
reconstruction is $G\simeq 0.998$. Using the estimated value of
$\sigma^2$ a deviation of about $13\%$ of the laser photon number
distribution from the Poissonian statistics can be derived.
\par
A second measurement was performed on the laser pulse diffused by
a moving ground glass. When the photons are collected from within
an area of spatial coherence, the system acts as a pseudo-thermal
source, whose photon number distribution satisfies
\begin{equation}
\varrho_{n,T} = \frac{N^n}{(N+1)^{n+1}}\: . \label{therm}\:
\end{equation}
Figure \ref{f:total} b) shows the photon distribution $\varrho_n$,
as reconstructed at the $i=400$ iteration and the best fit of the
data with (\ref{therm}) ($N= 5.33$); the fidelity is given by
$G\simeq 0.995$. The inset of the figure contains the experimental
frequency $f_{\nu}$ data ($K=24$ values of $\eta$, $n_\nu=10^4$
runs for each $\eta$) and their theoretical values as calculated
from (\ref{p0}).
\par
The last measurement was performed on the blue portion (420 nm) of
the down conversion fluorescence produced by a type I BBO crystal
(10 mm depth, cut at 34 deg) pumped by the laser pulse. The pump,
incident orthogonally to the crystal face, had an intensity $\sim
60$ GW/cm$^2$. In this experimental condition we expect a
coherence time of the generated field of $\sim 1$ ps, that
corresponds to measuring a convolution of 4-5 temporal modes
\cite{multit}. The photon number distribution is expected to be a
"multithermal" distribution of the form
\begin{equation}
\varrho_{n,M} = \frac{(n+\mu-1)!}{n!(\mu-1)!
(1+N/\mu)^n(1+\mu/N)^{\mu}}\: , \label{multi}\:
\end{equation}
where $\mu$ is the number of temporal modes. The photon
distribution reconstructed at the $i=1500$ iteration, is shown in
Fig. \ref{f:total} c) along with the best fit of the data using
(\ref{multi}) ($N= 6.17$ and $\mu = 5$); the fidelity of the
reconstruction is given by $G\simeq 1$. In the inset of the figure
we show the experimental frequency $f_{\nu}$ data ($K=18$ values
of $\eta$, $n_\nu=10^4$ runs for each $\eta$) and their
theoretical values as calculated according to (\ref{p0}). As a
comment to the experimental results in the pulsed regime, we note
that the best reconstruction of the photon distribution is
achieved at a different number of iterations for the three
different measured optical states, and that the absolute error
$\varepsilon$ does not approach the same value. This is due to the
presence of excess noise in our measurements, since the stability
and the repetition rate of our source (500 Hz) limits to
$n_\nu\sim 10^4$ the number of runs for each value of the quantum
efficiency \cite{pcount}. The choice of the best iteration to stop
the algorithm is driven by the possibility to fit the distribution
with a suitable model. We stress that there was no \emph{a-priori}
decision in choosing a Gaussian distribution for case a) or of a
multithermal distribution for case c), but, on the contrary, we
followed the \emph{a-posteriori} observation that no other
distribution could fit equally well the reconstructed data.
\section{Conclusions}
In summary, we have presented experimental results on the
reconstruction of the photon distribution based on on/off
detection at different quantum efficiency followed by a
maximum-likelihood iterative algorithm based on the theoretical
analysis presented in \cite{pcount}.
\par
Our results concern single-photon (PDC heralded) and weak coherent
states in the cw regime, as well as to coherent, thermal and
multi-thermal states in the pulsed regime. The stability and the
good accuracy shown in the reconstruction of these states,
together with the simplicity of the method, clearly demonstrate
the interesting potentialities of this technique, suggesting
relevant future applications ranging from studies on quantum
optics,  foundations of quantum mechanics, quantum information and
quantum metrology.
 Some applications in these fields are now
under realisation in our laboratories (reconstruction of further
quantum optical states, of entangled states \cite{en},
characterization of a high spectral selection heralded photon
source \cite{minsk}, etc.).
\section{Acknowledgement}
This work has been supported by MIUR (FIRB RBAU01L5AZ-002 and
RBAU014CLC-002), by INFM (PRA-CLON), by "Regione Piemonte" and by
Fondazione San Paolo. AR and MGAP thanks Stefano Olivares for many
fruitful discussions.


\end{document}